\documentclass[%
 amsmath,amssymb,
 aps,
 twocolumn,
prb,
floatfix,
citeautoscript,
10pt
]{revtex4-2}

\setcitestyle{super}

\usepackage[caption=false]{subfig}

\usepackage{graphicx}% Include figure files
\usepackage{dcolumn}% Align table columns on decimal point
\usepackage{bm}% bold math
\usepackage{longtable}
\usepackage{hyperref} % add hypertext capabilities
\usepackage[version=3]{mhchem}
\usepackage{color}

\newcolumntype{C}[1]{>{\centering\arraybackslash}m{#1}}
\usepackage[normalem]{ulem}
\usepackage{physics}
\usepackage{enumitem}
\usepackage{siunitx}

\begin{document}

\preprint{APS/123-QED}

\title{Qubit-efficient quantum chemistry with the ADAPT variational quantum eigensolver and double unitary downfolding}
\author{Harjeet Singh}
\thanks{These authors contributed equally to this work}
\affiliation{Department of Chemistry, Virginia Tech, Blacksburg, VA 24061, USA}
\affiliation{Virginia Tech Center for Quantum Information Science and Engineering, Blacksburg, Virginia 24061, USA}

\author{Luke W. Bertels}
\email{bertelslw@ornl.gov}
\thanks{These authors contributed equally to this work}
\affiliation{Quantum Information Science Section, Oak Ridge National Laboratory, Oak Ridge, TN 37831, USA}

\author{Daniel Claudino}
\thanks{This manuscript has been authored by UT-Battelle, LLC,under Contract DE-AC0500OR22725 with the U.S. Department of Energy. The United States Government retains and the publisher, by accepting the article for publication, acknowledges that the United States Government retains a nonexclusive, paid-up, irrevocable, worldwide license to publish or reproduce the published form of this manuscript, or allow others to do so, for the United States Government purposes. The Department of Energy will provide public access to these results of federally sponsored research in accordance with the DOE Public Access Plan.}
\affiliation{Quantum Information Science Section, Oak Ridge National Laboratory, Oak Ridge, TN 37831, USA}

\author{Sophia E. Economou}
\author{Edwin Barnes}
\affiliation{Department of Physics, Virginia Tech, Blacksburg, VA 24061, USA}
\affiliation{Virginia Tech Center for Quantum Information Science and Engineering, Blacksburg, Virginia 24061, USA}
\author{Nicholas J. Mayhall}
\email{nmayhall@vt.edu}
\affiliation{Department of Chemistry, Virginia Tech, Blacksburg, VA 24061, USA}
\affiliation{Virginia Tech Center for Quantum Information Science and Engineering, Blacksburg, Virginia 24061, USA}

\author{Nicholas P. Bauman}
\affiliation{Physical Sciences and Computational Division, Pacific Northwest National Laboratory, Richland, WA 99354, USA}
\author{Karol Kowalski}
\affiliation{Physical Sciences and Computational Division, Pacific Northwest National Laboratory, Richland, WA 99354, USA}
\affiliation{
  Department of Physics, University of Washington, Seattle, Washington 98195, USA
 }

\date{\today}% It is always \today, today,

\begin{abstract}
In this work, we combine the recently developed double unitary coupled cluster (DUCC) theory  with the adaptive, problem-tailored variational quantum eigensolver (ADAPT-VQE) to explore accuracy of unitary downfolded Hamiltonians for quantum simulation of chemistry. 
We benchmark the ability of DUCC effective Hamiltonians to recover dynamical correlation energy outside of an active space.   
We consider the effects of strong correlation, commutator truncation, higher-body terms, and approximate external amplitudes on the accuracy of these effective Hamiltonians. 
When combining these DUCC Hamiltonians with ADAPT-VQE, we observe similar convergence of the ground state as compared to bare active space Hamiltonians, demonstrating that DUCC Hamiltonians provide increased accuracy without increasing the load on the quantum processor. 
\end{abstract}

\maketitle

%\tableofcontents

\section{\label{sec:intro}Introduction}
The accurate simulation of the ground states of molecular electronic systems lies at the heart of theoretical chemistry and material science but is complicated by the exponential growth of the corresponding Hilbert spaces. 
Classical algorithms utilizing low-rank approximate ans{\"a}tze often fail to capture the physics required to accurately describe systems where the electrons are strongly correlated.  
As a result, simulation of quantum many-body physics has been heralded as a promising use case for the developing technology of quantum computing,\cite{feynman1982simulating,bartschi2024potential}
where the quantum processor is more naturally equipped to simulate the exponentially growth of the corresponding Hilbert space\cite{aspuru2005simulated}.

The current state of quantum computation, often referred to as the noisy, intermediate-scale quantum (NISQ) era\cite{preskill2018quantum}, is characterized by hardware platforms with tens to thousands of physical qubits\cite{gambetta2022expanding,evered2023high}, a lack of robust error correction\cite{shor1995scheme,google2023suppressing,acharya2024quantum}, short coherence times\cite{wack2021quality}, imperfect gate controls\cite{underwood2021gate,kandala2021demonstration}, and limited qubit connectivity\cite{ogorman2019generalized}. 
These limitations of current generation quantum processors have motivated algorithmic strategies to best utilize these emerging computational platforms. 
Current research on quantum algorithms has focused on variational quantum algorithms\cite{peruzzo2014variational,cerezo2021variational} over quantum phase estimation\cite{abrams1997simulation,abrams1999quantum} due to the long circuit requirements of the latter. 

Variational quantum eigensolvers (VQEs) are a class of hybrid quantum-classical algorithms that employ a feedback loop between quantum and classical resources to approximate the ground state of a Hamiltonian\cite{peruzzo2014variational,mcclean2016theory}.
A parameterized trial state $|\psi(\vec{\theta})\rangle$ is prepared on the quantum processor via a parameterized quantum circuit corresponding to a unitary operator acting on a reference state: 
\begin{equation}
    |\psi(\vec{\theta})\rangle = \hat{U}(\vec{\theta})|\phi_{0}\rangle.
\end{equation}
This trial state is then used to measure the expectation value of the problem Hamiltonian (after transformation into the qubit basis), which is lower-bounded by the exact ground state energy,
a condition which allows the parameters to be defined via direct minimization: 
\begin{align}
    E &= \min_{\vec{\theta}} \langle\psi(\vec{\theta})|\hat{H}|\psi(\vec{\theta})\rangle \label{eq:variational_principle}\\
    &= \min_{\vec{\theta}} \sum_{i} h_{i} \langle \psi(\vec{\theta}) | \hat{o}_{i} | \psi(\vec{\theta}) \rangle\label{eq:pauli_words},
\end{align}
where $\hat{o}_{i}$ are the component Pauli strings and $h_{i}$ are the corresponding weights. 
These expectation values are statistically converged over many repetitions of state preparation and measurements (shots), and as different Pauli strings $\hat{o}_i$ in Eq.~\ref{eq:pauli_words} do not generally commute naively they must be measured and converged separately. 
Recent strategies for reducing the shot count overhead of VQEs include efficient grouped measurement schemes\cite{ogorman2019generalized,mcclean2016theory,kandala2017hardware,rubin2018application,gokhale2019minimizing,jena2019pauli,izmaylov2019unitary,izmaylov2019revising,verteletskyi2020measurement,yen2020measuring,huggins2021efficient}, state tomography via classical shadows\cite{huang2020predicting,zhao2021fermionic}, positive operator-valued measurements\cite{garcia2021learning}, variance minimization\cite{wecker2015progress,yen2023deterministic,arrasmith2020operator,crawford2021efficient,zhang2023composite,mniszewski2021reduction,zhu2024optimizing}, and machine learning\cite{liang2024artificial}.   
Once the expectation values have converged the classical co-processor is then used to update the ansatz parameters $\vec{\theta}$ for the next iteration. 

Several ans\"atze have been explored in the context of VQE, generally falling into the categories of ``chemistry-/physics-inspired"\cite{peruzzo2014variational,mcclean2016theory,omalley2016scalable,barkoutsos2018quantum,romero2018strategies,colless2018computation,lee2018generalized,shen2017quantum,zhao2023orbital,mazzola2019nonunitary,matsuzawa2020jastrow,motta2023bridging,robledo2024chemistry}, usually a modification of unitary coupled cluster (UCC)\cite{bartlett1989alternative,kutzelnigg1991error} or Jastrow\cite{jastrow1955many} ans\"atze from classical electronic structure theory, or ``hardware-efficient"\cite{kandala2017hardware,ryabinkin2018qubit}, which form an ansatz by applying alternating layers of single-qubit rotations and entangling gates. 
While an arbitrarily expressive circuit could, in principle, provide a mapping between any two states in Hilbert space (including from a reference to the exact ground state), the realities of NISQ-era computing impose practical limits on the unitary ansatz employed. 
As a consequence, VQEs generally provide approximations to the ground state with accuracy defined by the variational flexibility of the ansatz. 

% adaptive algorithms (ADAPT-VQE, VQITE, Qubit CC, projective quantum eigensolver)

Adaptive VQEs are a class of quantum algorithms that seek to iteratively construct more accurate ans\"atze than VQEs with predefined ans\"atze\cite{grimsley2019adaptive,ryabinkin2020iterative,stair2021simulating,smart2021quantum,smart2022resolving,boyn2021quantum,burton2023exact}. 
The adaptive problem-tailored VQE (ADAPT-VQE) of Grimsely et al.\cite{grimsley2019adaptive} performs this ansatz construction by appending exponentiated operators from an operator pool to a trial state using the commutator of the pool operators and the problem Hamiltonian as a heuristic for selection. 
This more flexible and problem-informed ansatz has found solutions with both lower gate counts and lower errors than fixed-ansatz VQEs\cite{grimsley2019adaptive,claudino2020benchmarking,bertels2022symmetry}.
While initially formulated in the context of trotterized UCC operators, ADAPT-VQE has been generalized to use operator pools of Pauli strings (qubit-ADAPT-VQE)\cite{tang2021qubit}, qubit excitations (QEB-ADAPT-VQE)\cite{yordanov2021qubit} and coupled exchange operator (CEO-ADAPT-VQE)\cite{mafalda2024reducing}. 
%\ed{You could also mention CEO-ADAPT-VQE (https://arxiv.org/abs/2407.08696) since that performs even better.}  
Other extensions and inspirations of ADAPT-VQE include adaptive variational quantum imaginary time evolution approaches (AVQITE)\cite{gomes2021adaptive}, adaptive approaches to the quantum approximate optimization algorithm (QAOA)\cite{zhu2022adaptive}, acceleration using overlap with classical wavefunctions\cite{feniou2023overlap}, and dense tiling of ADAPT-VQE circuits (TETRIS-ADAPT-VQE)\cite{anastasiou2024tetris}.

% How to add dynamical correlation to VQEs
While many quantum algorithms seek to address the strong (static) correlation problem, simulations of chemical systems require large orbital basis sets to capture weak (dynamical) correlation arising from instantaneous motions of the electrons and produce experimentally-comparable results.  
Due to limited numbers of qubits on near-term devices, however, full encoding of this large number of spin-orbitals lies out of reach. 
A promising way towards demonstrating utility in the NISQ era could incorporate classical approaches to reduce the qubit costs of quantum algorithms\cite{bauer2016hybrid}.

A classical strategy for simplifying complex molecular systems reduces the problem onto an active space of important configurations. 
Takeshita et al.\cite{takeshita2020increasing} included orbital optimization with an active space approach to quantum subspace expansion in analogy to the classical multi-configurational self-consistent field method\cite{szalay2012multiconfiguration}. 
Following this, several studies have included orbital optimization into hybrid quantum-classical algorithms, including ground-state VQE\cite{mizukami2020orbital,sokolov2020quantum,tilly2021reduced,bierman2023improving,degracia2023complete,zhao2023orbital}, excited-state VQE\cite{yalouz2021state,gocho2023excited,omiya2022analytical}, and adaptive VQE\cite{fitzpatrick2024self}.  
Another class of methods apply post-VQE corrections using the results of an active space VQE. 
Such post-processing methods include 
%the anti-Hermitian contracted Schr{\"o}dinger equation (ACSE)\cite{smart2021quantum,smart2022resolving,boyn2021quantum}, 
multiconfigurational pair density functional theory\cite{boyn2021quantum}, symmetry-adapted perturbation theory (SAPT)\cite{malone2022towards,loipersberger2023accurate}, and $N$-electron valence perturbation theory (NEVPT)\cite{tammaro2023n}. 
Other perturbative approaches have been employed to improve the accuracy of unitary coupled-cluster-based VQE\cite{windom2024attractive,windom2024new} and iterative qubit coupled-cluster VQE\cite{ryabinkin2021posteriori}. 
Embedding techniques offer another approach from classical theory to extend the reach of quantum simulations towards larger systems.
Density matrix embedding theory (DMET) has been applied to the study of hydrogen chains\cite{kawashima2021optimizing} and model hydrocarbons\cite{li2022toward}. 
Quantum defect embedding theory has been used to study spin defects with VQE and quantum subspace expansion\cite{huang2022simulating}. 
Projection-based embedding has been utilized to study small molecules with a VQE-in-DFT approach\cite{rossmannek2023quantum}. 

Effective Hamiltonians from a downfolding transformation offer an alternative approach to including dynamical correlation effects into an active space. 
The effective Hamiltonians are generated by decoupling the active and inactive spaces via a transformation and projection of the full system Hamiltonian. 
This transformation then approximates the correlation effects of the inactive space within the active space. 
Different formulations for Hamiltonian downfolding for quantum algorithms have arisen out of explicit correlation theory\cite{motta2020quantum,mcardle2020improving,kumar2022quantum,schleich2022improving,dobrautz2024toward,magnusson2024towards}, Schrieffer-Wolff transformation\cite{karol2023fock,zhang2022quantum}, driven similarity renormalization group theory\cite{huang2023leveraging}, and coupled-cluster theory\cite{kowalski2018properties,kowalski2023sub,bauman2019downfolding,bauman2019quantum,bauman2020coupled,kowalski2020sub,metcalf2020resource,chladek2021variational,bauman2022coupled,bauman2022double,bauman2023coupled}. 

Double unitary coupled cluster theory (DUCC) provides a single reference formulation for constructing active space downfolded Hamiltonians by partitioning the cluster amplitudes into those internal and external to the active space\cite{bauman2019downfolding}. 
The effective Hamiltonian is then formed by similarity transforming the Hamiltonian with the external amplitudes and projecting onto the active space. 
The unitary formulation results in Hermitian effective Hamiltonians which are more naturally suited to simulation with quantum hardware. 
The DUCC procedure has been used in conjunction with VQE and quantum subspace expansion to study the electronic structure of small molecules on simulated quantum hardware\cite{metcalf2020resource}. 
The quantum flow algorithm (QFlow) also utilized DUCC downfolding to treat quantum many-body systems by self-consistently optimizing a global pool of amplitudes\cite{kowalski2023quantum}. 

In this work, we investigate the accuracy of DUCC effective Hamiltonians in combination with ADAPT-VQE. 
We explore the accuracy of these Hamiltonians across different correlation regimes, truncations of the commutator expansion, higher-body effects, and external amplitudes used in their construction. 
We also investigate the convergence of ADAPT-VQE in the presence of these transformed integrals. 
%In this work, we investigate the performance of ADAPT-VQE in combination with DUCC downfolded Hamiltonians for finding accurate ground states of small molecular systems on an ideal simulator. 
%We benchmark the performance of ADAPT-VQE-DUCC across correlation regimes, active spaces, and truncations of the effective Hamiltonians. 
%We also explore the accuracy of using second-order M{\o}ller-Plesset (MP2)\cite{moller1934note} amplitudes to approximate the external cluster amplitudes in the downfolding procedure. 

This work is organized as follows: Section~\ref{sec:background} provides a background of the ADAPT-VQE and DUCC algorithms, Section~\ref{sec:details} outlines the computational details of the simulations, and Section~\ref{sec:results} presents and discusses the performance of DUCC-ADAPT-VQE for several model systems. 

\section{\label{sec:background}Background}
\subsection{\label{ssec:ADAPT}ADAPT-VQE Algorithm}
The ADAPT-VQE algorithm iteratively constructs a parameterized unitary for VQE by concatenating exponentiated operators from a pool as informed by the problem Hamiltonian\cite{grimsley2019adaptive}. 
At the start of the algorithm, the user defines a spin-orbital to qubit mapping, operator pool, and reference state.
The operators $\{\hat{A}_{k}\}$ used in this work are the generators of the trotterized, generalized unitary coupled cluster with single and double excitations/de-excitations:
\begin{align}
\hat{A}^{p}_{q} &= \hat{a}^{\dagger}_{p}\hat{a}_{q} - \hat{a}^{\dagger}_{q}\hat{a}_{p} \label{eq:singles}\\
\hat{A}^{pq}_{rs} &= \hat{a}^{\dagger}_{p}\hat{a}^{\dagger}_{q}\hat{a}_{s}\hat{a}_{r} - \hat{a}^{\dagger}_{r}\hat{a}^{\dagger}_{s}\hat{a}_{q}\hat{a}_{p} \label{eq:doubles},
\end{align}
where $\hat{a}^{\dagger}$ and $\hat{a}$ are fermionic creation and annihilation operators and $p$, $q$, $r$, and $s$ are general spin-orbital indices. 
The operator pool may be spin-complemented or spin-adapted by taking linear combinations of these anti-Hermitian generators (see Appendix of Ref.~\citenum{bertels2022symmetry}). 
Different operator pools have been explored in other studies\cite{tang2021qubit,yordanov2020efficient,yordanov2021qubit,shkolnikov2023avoiding}. 
The reference state, an easily preparable state on the device (usually a product state mapping to the Hartree-Fock determinant), is used to initialize the ADAPT-VQE trial state. 

To begin an ADAPT-VQE iteration, the current trial state, $|\psi^{(n)}\rangle$ is prepared and used to measure the expectation value of the operator gradient:
\begin{equation}
    \frac{\partial E^{(n)}}{\partial\theta_{k}} = \left\langle \psi^{(n)} \left|[\hat{H},\hat{A}_{k}]\right| \psi^{(n)} \right\rangle. \label{eq:op_grad}
\end{equation}
This step formally requires measurement of the three-body reduced density matrix (3-RDM) but is parallelizable over multiple quantum devices. 
The ansatz is grown by exponentiating and appending the operator corresponding to the largest magnitude gradient to the previous trial state:
\begin{align}
    |\psi^{(n+1)}(\vec{\theta}^{(n+1)})\rangle &= e^{\theta_{n+1}\hat{A}_{n+1}}|\psi^{(n)}\rangle \\
    &= e^{\theta_{n+1}\hat{A}_{n+1}}e^{\theta_{n}\hat{A}_{n}}\cdots e^{\theta_{1}\hat{A}_{1}}|\psi^{(0)}\rangle.
\end{align}
The parameters of this new ansatz are then optimized with a VQE subroutine, taking as initial parameters the optimized parameters from the previously optimized trial state and initializing the new parameter $\theta_{n+1}$ to zero. 
With this new trial state the algorithm iterates and returns to the operator gradient measurement step. 
Convergence of the algorithm is typically determined by a threshold on the operator gradient. 

The addition of an operator to the ansatz does not remove that operator from the pool, and as such the same operator may be added multiple times with independent parameters for each instance.
When combined with a complete operator pool, this property allows ADAPT-VQE to approach exact eigenstates of the Hamiltonian\cite{evangelista2019exact,grimsley2019adaptive}. 

\subsection{\label{ssec:DUCC}Double Unitary Coupled Cluster Downfolding}
The DUCC formulation\cite{bauman2019downfolding} is a unitary adaptation of the subsystem embedding subalgebras coupled-cluster (SES-CC) method\cite{kowalski2018properties} to yield Hermitian effective Hamiltonians. 
The ansatz underlying DUCC takes a double unitary form: 
\begin{equation}
    |\Psi\rangle = e^{\hat{\sigma}_{\text{ext}}}e^{\hat{\sigma}_{\text{int}}}|\phi_{0}\rangle,
\end{equation}
where $|\phi_{0}\rangle$ is the single reference and $\hat{\sigma}_{\text{ext}}$ and $\hat{\sigma}_{\text{int}}$ are the anti-Hermitian external and internal cluster generators, respectively:
\begin{align}
    \hat{\sigma}_{\text{ext}}^{\dagger} & = -\hat{\sigma}_{\text{ext}}, \\
    \hat{\sigma}_{\text{int}}^{\dagger} & = -\hat{\sigma}_{\text{int}}.
\end{align}
These generators are defined such that the internal generators act only on spin-orbitals within the active space and the external generators act on at least one orbital outside of the active space. 

With this definition for the DUCC ansatz, the energy is determined via a Hermitian eigenvalue problem:
\begin{equation}
    \hat{H}_{\text{eff}} e^{\hat{\sigma}_{\text{int}}}|\phi_{0}\rangle = Ee^{\hat{\sigma}_{\text{int}}}|\phi_{0}\rangle,
\end{equation}
where the effective Hamiltonian in the active space is given by
\begin{equation}
    \hat{H}_{\text{eff}} = (\hat{P}+\hat{Q}_{\text{int}})\hat{H}_{\text{ext}}(\hat{P}+\hat{Q}_{\text{int}}),
\end{equation}
%\ed{Define $\hat{P}$ and $\hat{Q}_\mathrm{int}$.}
$\hat{P}$ and $\hat{Q}_{\text{int}}$ are the projectors onto the reference state and excited configurations within the active space, respectively, and 
\begin{equation}
    \hat{H}_{\text{ext}} = e^{-\hat{\sigma}_{\text{ext}}}\hat{H}e^{\hat{\sigma}_{\text{ext}}}. \label{eq:sim_trans}
\end{equation}
This effective Hamiltonian may then be diagonalized with quantum or classical active space solvers. %The $\hat{Q}_\mathrm{int}$ operator projects onto the excited configurations within the active space relative to $|\phi_{0}\rangle$, while the projection onto the reference function is represented by $\hat{P}$.

When constructing the second-quantized representation of $\hat{H}_{\text{eff}}$, we begin by transforming the bare Hamiltonian into its particle-hole normal ordered form by subtracting off the reference energy,
\begin{equation}
    \hat{H}_{N} = \hat{H} - \langle \phi_{0} | \hat{H} | \phi_{0} \rangle,
\end{equation}
which can further be separated into its one- and two-body components,
\begin{equation}
    \hat{H}_{N} = \hat{F}_{N}+\hat{V}_{N}.
\end{equation} 
Because the components of $\hat{\sigma}_{\text{ext}}$ do not generally commute, the Baker-Campbell-Hausdorff (BCH) expansion of Eq.~\eqref{eq:sim_trans} does not terminate at finite order and must be truncated. 
In this work, we consider two truncations of $\hat{H}_{\text{ext}}$:
\begin{align}
    \hat{H}_{\text{ext}}^{\text{A4}} =& \hat{H}_{N} + [\hat{H}_{N},\hat{\sigma}_{\text{ext}}] + \frac{1}{2}[[\hat{F}_{N},\hat{\sigma}_{\text{ext}}],\hat{\sigma}_{\text{ext}}], \label{eqn:a4}\\
    \hat{H}_{\text{ext}}^{\text{A7}} =& \hat{H}_{N} + [\hat{H}_{N},\hat{\sigma}_{\text{ext}}] + \frac{1}{2}[[\hat{H}_{N},\hat{\sigma}_{\text{ext}}],\hat{\sigma}_{\text{ext}}] \label{eqn:a7} \\
    & + \frac{1}{6}[[[\hat{F}_{N},\hat{\sigma}_{\text{ext}}],\hat{\sigma}_{\text{ext}}],\hat{\sigma}_{\text{ext}}], \nonumber
\end{align}
that are correct through second and third order, respectively then truncated to two body operators.
We use the original labels defining these truncations from Ref.~\citenum{bauman2022double}. 
Although higher-body terms arise from these commutator expansions, previous studies have only retained the one- and two-body interactions in $\hat{H}_{\text{ext}}$. 
In this work, we investigate the role of these higher-body interactions in the effective Hamiltonians. 
While the external amplitudes are typically approximated from projected coupled cluster theory with single and double excitations (CCSD)\cite{purvis1982full},  
\begin{equation}
    \hat{\sigma}_{\text{ext}} \approx \hat{T}^{\text{CCSD}}_{1,\text{ext}} + \hat{T}^{\text{CCSD}}_{2,\text{ext}} - (\hat{T}^{\text{CCSD}}_{1,\text{ext}})^{\dagger} - (\hat{T}^{\text{CCSD}}_{2,\text{ext}})^{\dagger},
\end{equation}
we also use MP2\cite{moller1934note} amplitudes to approximate the external amplitudes:
\begin{equation}
    \hat{\sigma}_{\text{ext}} \approx \hat{T}^{\text{MP2}}_{2,\text{ext}} - (\hat{T}^{\text{MP2}}_{2,\text{ext}})^{\dagger}.
\end{equation}
We additionally explore the use of both CCD amplitudes and the doubles amplitudes from a CCSD calculation to form the approximate external amplitudes:
\begin{align}
    \hat{\sigma}_{\text{ext}} &\approx \hat{T}^{\text{CCD}}_{2,\text{ext}} -  (\hat{T}^{\text{CCD}}_{2,\text{ext}})^{\dagger}, \\
    \hat{\sigma}_{\text{ext}} &\approx \hat{T}^{\text{CCSD}}_{2,\text{ext}} -  (\hat{T}^{\text{CCSD}}_{2,\text{ext}})^{\dagger}.
\end{align}
\section{\label{sec:details}Computational Details}
The PySCF software package\cite{sun2018pyscf, sun2020recent} was used to generate integrals in the canonical orbital basis. 
Natural virtual orbitals were then obtained by diagonalizing the virtual space of the one-particle reduced density matrix (1-RDM) from an MP2 calculation to define the active space in the cc-pVTZ basis\cite{dunning1989gaussian,prascher2011gaussian}. 
The Wick\&d package\cite{evangelista2022wicked} was used to generate commutator expressions for the BCH expansion, followed by the use of OpenFermion\cite{mcclean2020openfermion} to transform the effective Hamiltonian via the Jordan-Wigner transformation for ADAPT-VQE calculations with an in-house code. 
The generalized singles and doubles operator pool was chosen for the ADAPT-VQE calculations. 
In this work, we study three benchmark systems: \ce{LiH}, \ce{H6}, and \ce{H2O}. 
For \ce{H6}, we consider linear geometries with equal separation between adjacent hydrogen atoms.
For \ce{H2O} single bond dissociation, we use an equilibrium \ce{O-H} bond length of 0.96183 \AA ~and a bond angle of 103.9215\textdegree. 
Active spaces were selected by analyzing natural orbital occupations: 8 orbitals for \ce{LiH}, 6 for \ce{H6}, and 9 for \ce{H2O}. 
The errors for the \ce{LiH} system (44 orbitals) are calculated relative to full configuration interaction (FCI) calculations, while the errors for the \ce{H6} (84 orbitals) and \ce{H2O} (58 orbitals) systems were calculated with respect to the extrapolated Adaptive Sampling CI (ASCI) calculations, performed using the MACIS package\cite{williams2023asci}. 
To obtain the extrapolated ASCI energy, three ASCI calculations were performed using 1 million, 5 million, and 10 million determinants, and these values were then extrapolated to obtain the ASCI energy in the zero PT2 limit.

\section{\label{sec:results}Results}
 We present the results of our analysis in three parts. 
In the first part, we examine the performance of DUCC Hamiltonians in accurately capturing external correlation energies. 
This analysis is further divided into two subsections: The first focuses on bond breaking, while the second explores the significance of three- and higher-body terms in the construction of DUCC effective Hamiltonians. 
The second part investigates alternative approximations for the DUCC external amplitudes. 
Finally, the third part studies the convergence behavior of the ADAPT-VQE algorithm with the DUCC Hamiltonians. 

\subsection{Accuracy of DUCC Effective Hamiltonians}
\subsubsection{Potential Energy Surface Scans}
\begin{figure*}[htp]
    \centering
    \includegraphics[width=\textwidth]{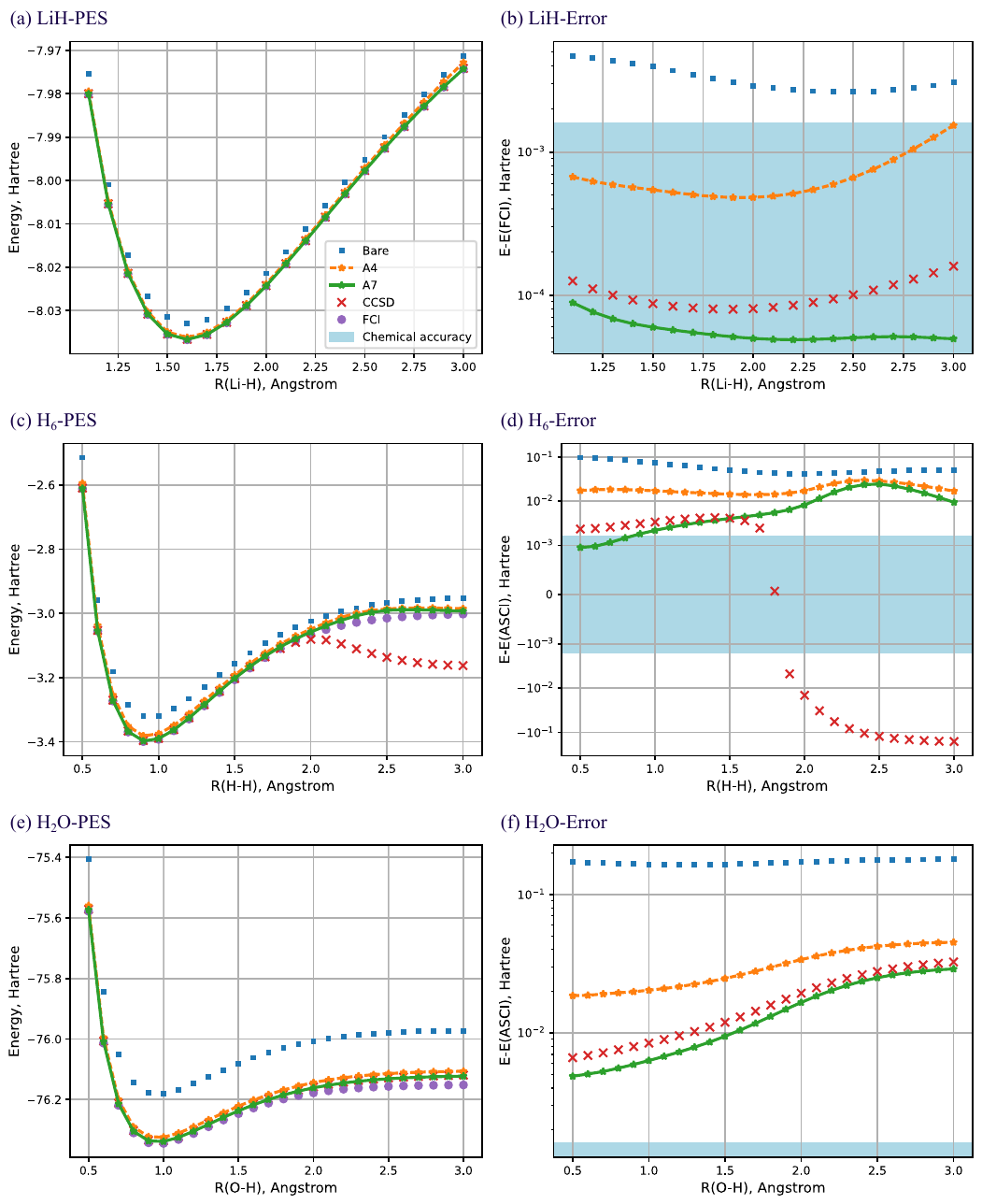}
    \caption{Comparison of the ground states of the bare (blue squares), A4 (orange stars, dashed line), and A7 (green stars, solid line) Hamiltonians for \ce{LiH} dissociation (8 orbital active space), symmetric linear \ce{H6} dissociation (6 orbital active space), and \ce{H2O} single-bond dissociation (9 orbital active space). The left panel shows the absolute energies, while the right panel shows corresponding errors. The error for \ce{LiH} is calculated relative to the FCI energy (violet dots), while the errors for \ce{H6} and \ce{H2O} are calculated relative to the extrapolated ASCI energy(violet dots). Energies and errors for CCSD in the full orbital space are presented in red. The blue shaded region represents a chemical accuracy of 1kcal/mol (1.59 mHa).}
    \label{fig:pes_lih_h6_h2o}
\end{figure*}

To evaluate the performance of DUCC downfolding, we compare the accuracy of three exact diagonalization (ED) ground state potential energy surface (PES) scans using the A4 and A7 approximations (Eqs.~\eqref{eqn:a4} and \eqref{eqn:a7}, respectively). 
The results of these calculations are presented in Fig.~\ref{fig:pes_lih_h6_h2o}, where the left panels show absolute energies and the right panels show the corresponding errors. 

Figures~\ref{fig:pes_lih_h6_h2o}(a) and \ref{fig:pes_lih_h6_h2o}(b) show the PES's and errors from FCI for the dissociation of \ce{LiH} in an 8 orbital active space.  
Exact diagonalization of the bare Hamiltonian in the active space fails to achieve chemical accuracy across the internuclear distances sampled, with an average error from FCI of 3.29 mHa and a non-parallelity error (NPE) of 2.07 mHa.  
Upon inclusion of external dynamical correlation approximated using CCSD amplitudes, the A4 approximation achieves a chemically accurate ground state throughout the PES. 
The A7 Hamiltonian performs even better due to the inclusion of higher-order terms in the commutator expansion, yielding an average error from FCI of 0.06 mH and an NPE of 0.04 mH. 
For this system, both CCSD and the DUCC effective Hamiltonians perform well, yielding chemically accurate results across the PES.
Additionally, the A7 approximation consistently outperforms CCSD across the PES. 

For the symmetric dissociation of linear \ce{H6}, Figs.~\ref{fig:pes_lih_h6_h2o}(c) and \ref{fig:pes_lih_h6_h2o}(d) show PES results using a 6 orbital active space. 
For this system, CCSD fails to provide an accurate description of the correlation energy as the bond distance is stretched, becoming non-variational for $R_{H-H}>1.8$ \AA. 
%For \ce{H6} (Figure \ref{fig:pes_lih_h6_h2o}(c) and \ref{fig:pes_lih_h6_h2o}(d)), the CCSD method fails, providing non-variational energies. 
The A4 and A7 methods are also seen to underestimate the correlation energy here, with A4 not achieving chemical accuracy and A7 losing chemical accuracy after 0.8 \AA. 
Comparing the active space methods here, the bare Hamiltonian yields an average error of 58.02 mH and an NPE of 55.17 mH, the A4 approximation yields an average error of 19.37 mH and an NPE of 15.68 mH, and the A7 approximation yields an average error of 8.84 mH and an NPE of 23.52 mH. 
%In contrast, DUCC with A4 and A7 approximations improves the active space energy and maintains chemical accuracy up to a bond distance of 0.8 Å. 
% Near 2.5 \AA, the DUCC approximations demonstrate very little improvement over the bare active space Hamiltonian.
Unlike CCSD, both the A4 and A7 approximations remain variational across the PES when using MP2 natural virtual orbitals. 
This is not necessarily the case for DUCC in general, as the truncation of the BCH expansion precludes a variational bound (see SI for examples). 
%Beyond this distance, as the system becomes more strongly correlated, DUCC continues to provide variational energies when using MP2 natural virtual orbitals. 
%However, it is important to note that canonical orbitals can sometimes produce non-variational energies. 
Therefore, it is recommended to use natural orbitals with DUCC, given that a sufficient number of orbitals are included in the active space to capture static correlation effects.
Overall, the DUCC approximations are seen to struggle when using this minimal active space for a strongly correlated system where CCSD breaks down. 
This breakdown of CCSD affects the DUCC approximations because of the use of CCSD amplitudes to build the external amplitudes.
The effect of these poor amplitudes is explored in the next section. 

For the \ce{H2O} single-bond-breaking system (Figs. \ref{fig:pes_lih_h6_h2o}(e) and \ref{fig:pes_lih_h6_h2o}(f)), the bare Hamiltonian in a 9 orbital active space exhibits errors greater than $10^{-1}$ Ha. 
Without the inclusion of external correlation effects or orbital relaxation, this truncation leads to significant errors relative to the other methods surveyed, highlighting the importance of including external correlations. 
The CCSD results in the full orbital space remain variational, suggesting that CCSD can provide a qualitatively correct description of the system; however, the CCSD results still show significant errors across the PES. 
DUCC with A4 and A7 approximations significantly improves the bare Hamiltonian energy by incorporating external dynamical correlation effects, although the results are not chemically accurate across the entire PES.
The A7 ground state consistently outperforms the CCSD result across the PES, while the A4 ground state roughly tracks the CCSD ground state. 

To probe the choice of active space, we investigated active space sizes from 9-13 orbitals for \ce{H2O} single bond dissociation at equilibrium and stretched geometries (see SI). 
From these calculations, we observe adding orbitals to the active space in this range does not meaningfully change the errors for either the A4 or A7 effective Hamiltonians. 
This would suggest that the CCSD amplitudes, while remaining variational in these cases, are poor approximations to the unitary external amplitudes, contributing to the failure of these methods to achieve chemical accuracy.

\subsubsection{Effect of higher-body terms in the expansion}
The nested commutator expansions presented in Eqs.~\eqref{eqn:a4} and \eqref{eqn:a7} include contributions that amount to three-body terms in the A4 approximation and three- and four-body terms in the A7 approximation, but these are typically neglected when forming the effective Hamiltonians. 
We denote the A4 and A7 approximations with three-body terms included as A4(3) and A7(3), respectively, and the A7 approximation with three- and four-body terms as A7(34). 
To analyze the significance of higher-body terms in DUCC effective Hamiltonians, we performed calculations across a range of systems: \ce{LiH} at $1R_e$ and $2R_e$ (with $R_e = 1.5949$ \AA), \ce{H6} at 1\r{A} and 2\r{A}, and \ce{H2O} single bond dissociation at $1R_e$ and $2R_e$ (with $R_e = 0.96183$ \AA ~and a bond angle of 103.9215\textdegree), where $R_e$ denotes equilibrium bond lengths. For each system, we examined the exact ground states of the bare, A4, A4(3), A7, A7(3), and A7(34) Hamiltonians and compared the results against SCF, MP2, and CCSD in the full orbital space. The results are shown in Table~\ref{tab:error_table}.

%The \ce{LiH} results demonstrate that the bare Hamiltonian, with an error of 3.73 mH at $1R_e$, systematically improves as higher-body terms are incorporated, achieving chemical accuracy with A7(34). 
%This trend is consistent at $2R_e$.
Beginning with the \ce{LiH} results, at both 1$R_{e}$ and $2R_{e}$ the shift in ground state energies when moving from A4 to A4(3) amounts to a few tens of $\mu$Ha. 
The shift when moving from A7 to A7(3) to A7(34) is of the same order of magnitude. 
These results are not altogether surprising as the A4 approximation and CCSD both perform quite well for these systems. 
Similarly, for \ce{H6}, while the bare Hamiltonian has substantial errors (73.46 mH at 1\r{A} and 41.84 mH at 2\r{A}), DUCC with A4 and A7 significantly reduces errors, particularly at larger bond distances where A7(3) and A7(34) provide variational energies and better agreement.
In \ce{H2O}, single-bond dissociation highlights the limitations of lower-order truncations. 
Errors in bare Hamiltonian calculations exceed 150 mH but decrease significantly with DUCC approximations. 
%For instance, the A4 approximation reduces the error to 32.27 mH at $2R_e$, while A7 and its extensions further improve accuracy. 
At $1R_e$, the A4 approximation lowers the error to 20.13 mH, while A7 and its extensions—A7(3) and A7(34)—all yield identical errors of 6.16 mH. At $2R_e$, the A4 approximation gives an error of 32.27 mH, while A7 improves this to 15.24 mH. The A7(3) and A7(34) approximations further decrease the error to 14.25 mH and 14.31 mH, respectively.

Notably, three- and four-body terms have a marginal impact on the correlation energy in weakly correlated systems but contribute significantly when $\hat{T}_1$ amplitudes are large.
This behavior is evident in the highly correlated \ce{H6} system at 2\r{A}. 
To support this observation, we performed the $\hat T_1$ diagnostic\cite{lee1989diagnostic} for \ce{H6}, obtaining values of 0.014 at 1\r{A} and 0.059 at 2\r{A} (0.010 and 0.058 for $\hat{T}_{1,\text{ext}}$, respectively),  %\ed{It's not clear to me what a $\hat T_1$ diagnostic is or what these numbers mean.}
indicating an increase in $\hat T_1$ amplitudes and correspondingly stronger correlations at larger bond distances.
%However, looking at the contributions from $\hat{T}_{2}$ we see comparable values for the external $\hat{T}_{2,\text{ext}}$ diagnostic at both 1 \AA and 2 \AA for \ce{H6}. 
When considering a $\hat{T}_{2}$ diagnostic, however, we observe that while the overall amplitude diagnostic increases when the \ce{H-H} separation is increased (0.130 to 0.450 for 1 and 2 \AA, respectively), the diagnostic for the $\hat{T}_{2,
\text{ext}}$ is similar for both geometries (0.071 and 0.058 for 1 and 2 \AA, respectively). 
As the three-body terms contributing to the A4(3) effective Hamiltonian only include contributions from commutators with $\hat{T}_{2,\text{ext}}$, this suggests that the magnitude of the three-body correction is not strictly tied to the magnitude of the external $\hat{T}_{2}$.
Overall, the data reveal that while higher-body terms in DUCC expansions contribute modestly to correlation energy in weakly correlated systems, they may become increasingly relevant in strongly correlated regimes where $\hat{T}_{1}$ amplitudes are larger, ensuring systematic energy improvements.
We report $\hat{T}_{1}$ and $\hat{T}_{2}$ diagnostics for these systems in the SI. 
\begin{table*}[ht]
    \caption{Errors (mHa) for SCF, MP2, and CCSD in the full basis and exact diagonalization of several active space Hamiltonians are presented with reference energies (Ha) for \ce{LiH} (8 orbital active space), linear \ce{H6} (6 orbital active space), and \ce{H2O} single bond dissociation (9 orbital active space). }
    \centering
    \renewcommand{\arraystretch}{1.2}
    \begin{tabular*}{\textwidth}{@{\extracolsep{\fill}} l r r r r r r}
        \hline\hline
        Method & \ce{LiH}(1$R_e$)$^a$ & \ce{LiH}(2$R_e$)$^a$ & \ce{H6}(1\AA)$^b$ & \ce{H6}(2\AA)$^b$ & \ce{H2O}(1$R_e$)$^b$ & \ce{H2O}(2$R_e$)$^b$\\ 
        \hline
        Reference & -8.03670 & -7.96733 & -3.39235 & -3.06639 & -76.34618 & -76.18447 \\
        SCF & 50.07 & 60.87 & 150.70 & 283.46 & 289.40 & 346.55 \\ 
        MP2 & 10.64 & 20.04 & 32.27 & 124.16 & 13.96 & 33.77 \\ 
        CCSD & 0.08 & 0.20 & 3.40 & -14.49 & 8.25 & 17.96 \\ 
        Bare & 3.73 & 3.49 & 73.46 & 41.84 & 165.18 & 170.27 \\ 
        A4 & 0.52 & 2.23 & 17.21 & 17.26 & 20.13 & 32.27 \\ 
        A4(3) & 0.52 & 2.14 & 17.15 & 11.48 & 20.12 & 31.11 \\ 
        A7 & 0.06 & 0.05 & 2.20 & 8.25 & 6.16 & 15.24 \\ 
        A7(3) & 0.05 & 0.09 & 2.11 & 3.19 & 6.16 & 14.25 \\ 
        A7(34) & 0.05 & 0.11 & 2.12 & 3.30 & 6.16 & 14.31 \\  
        \hline\hline
    \end{tabular*}
    \raggedright
    \footnotesize{$^a$ Error relative to FCI. $^b$ Error relative to ASCI. The FCI/ASCI energies are in Ha while remaining values are errors in mHa.}
    \label{tab:error_table}
\end{table*}

\subsection{Impact on DUCC energy with different types of amplitudes}
\begin{table}[ht]
    \caption{Error (mH) for \ce{H6} system at different bond lengths calculated with respect to ASCI.}
    \centering
    \renewcommand{\arraystretch}{1.2}
    \begin{tabular}{l l r r}
        \hline\hline
        Method &  & \ce{H6}(1\AA) & \ce{H6}(2\AA) \\ 
        \hline
        SCF &  & 150.70 & 283.46 \\ 
        MP2 &  & 32.27 & 124.16 \\ 
        CCSD &  & 3.40 & -14.49 \\ 
        Bare &  & 73.46 & 41.84 \\ 
        \hline
        MP2 Amps & A7 & 7.29 & 36.40 \\ 
                 & A7(3) & 7.30 & 36.25 \\ 
        \hline
        CCD Amps & A7 & 3.28 & 27.34 \\ 
                 & A7(3) & 3.21 & 23.16 \\ 
        \hline
        CCSD ($\hat T_1$=0) Amps & A7 & 3.27 & 27.22 \\ 
                            & A7(3) & 3.20 & 23.49 \\ 
        \hline
        CCSD Amps & A7 & 2.20 & 8.25 \\ 
                  & A7(3) & 2.11 & 3.19 \\ 
        \hline\hline
    \end{tabular}
    \label{tab:error_table_amps}
\end{table}
In this section, we investigate the impact of different types of amplitudes on the DUCC energy. Specifically, we tested the MP2, CCD, CCSD ($\hat T_1$=0), and CCSD amplitudes. The term ``CCSD ($\hat T_1$=0)'' refers to computing the CCSD amplitudes as usual and then setting \( \hat T_1 = 0 \), which allows for the \( \hat T_2 \) amplitudes to be solved in the presence of \( \hat T_1 \) amplitudes. For each set of amplitudes, we computed A7 and A7(3) DUCC Hamiltonians for the \ce{H6} system at bond lengths of 1\AA ~and 2\AA. The results are shown in Table~\ref{tab:error_table_amps}. 

From the table, we observe that CCSD provides a non-variational energy. When MP2 amplitudes are used with DUCC Hamiltonians, the error reduces from 73.46 mH (bare) to 7.29 mH (A7). Using CCD amplitudes, where the \( \hat T_2 \) amplitudes are optimized in the absence of \( \hat T_1 \), the error is further reduced to 3.28 mH. The CCSD ($\hat T_1$=0) amplitudes yield results similar to CCD, with only a slight difference. However, when CCSD amplitudes are used, the error decreases to 2.20 mH with the A7 Hamiltonian.

Notably, the error reduction from the bare Hamiltonian to the A7 Hamiltonian is significant across all amplitude types. The MP2 amplitudes show a reduction of approximately 66 mH at 1\AA ~and 5 mH at 2\AA, demonstrating that DUCC improves accuracy even with the simpler MP2 amplitudes. CCD and CCSD ($\hat T_1$=0) amplitudes provide similar improvements, with error reductions of around 70 mH at 1\AA ~and 14.5 mH at 2\AA. The most significant improvement occurs when using CCSD amplitudes, where the error drops to 2.2 mH at 1\AA ~and 8.25 mH at 2\AA, indicating the optimal performance of DUCC with these amplitudes.

Thus, we conclude that the DUCC energy is sensitive to the type of external amplitudes used. The best variational estimates are obtained when CCSD amplitudes are employed, as they improve the accuracy of the Hamiltonian significantly across different bond lengths.

\subsection{Convergence of ADAPT-VQE in the presence of transformed integrals}
\begin{figure*}[htp]
    \centering
    \includegraphics[width=\textwidth]{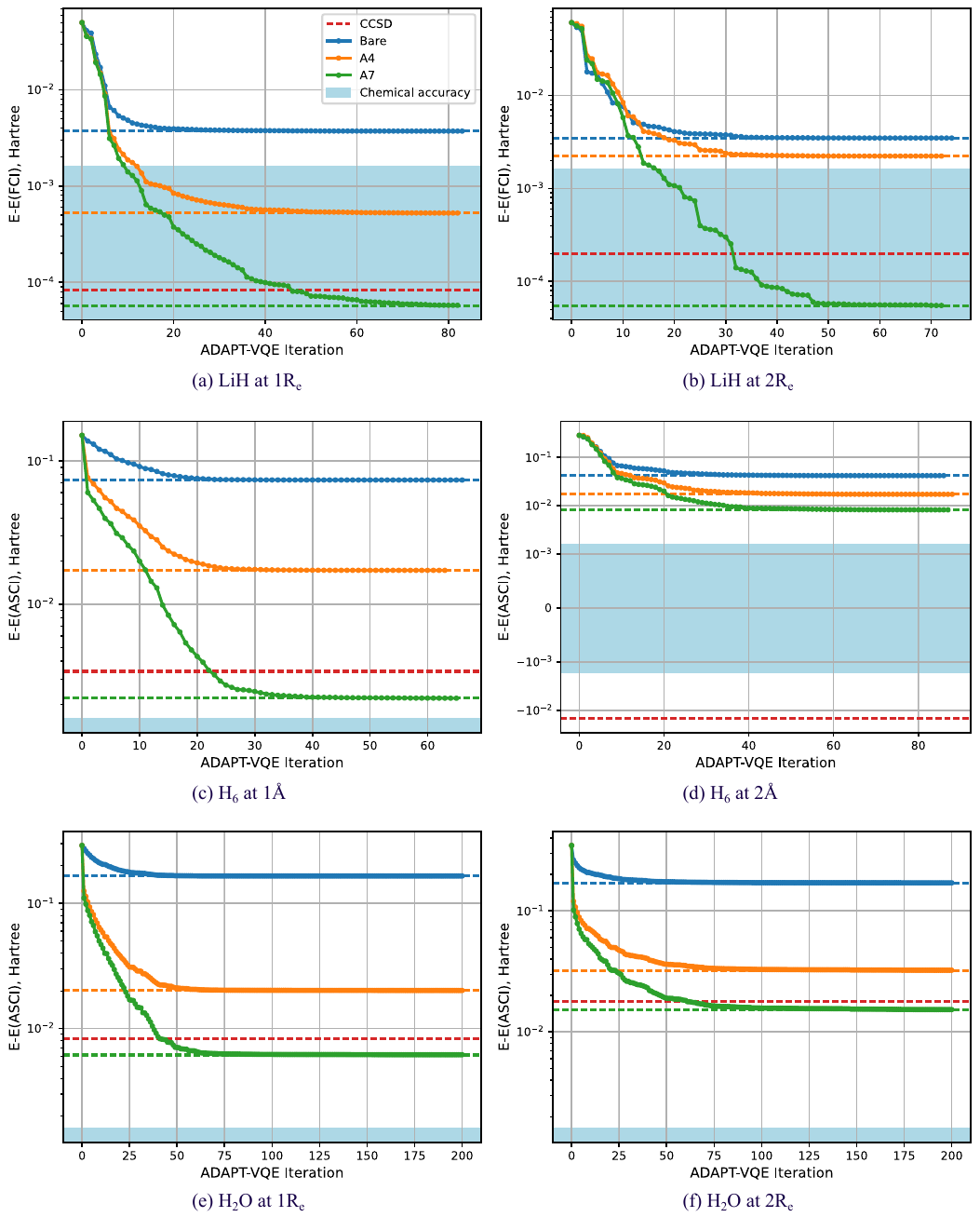}
    \caption{Energy error versus the number of ADAPT-VQE iteration for \ce{LiH} (8 orbital active space, 1$R_{\text{e}}$ and 2$R_{\text{e}}$ bond distance), \ce{H6} (6 orbital active space, 1\AA and 2\AA bond distance), and \ce{H2O} single bond dissociation (9 orbital active space, 1$R_{\text{e}}$ and 2$R_{\text{e}}$ bond distance). Results for full-space CCSD (red dashed line), bare (blue), A4 (orange), and A7 (green) Hamiltonians are shown. The blue, orange, and green dashed lines represent the exact diagonalization of the Hamiltonian in the active space. The blue shaded region indicates chemical accuracy (1 kcal/mol = 1.59 mHa).}
    \label{fig:adapt_lih_h6_h2o}
\end{figure*}

% To reduce the qubit requirements in ADAPT-VQE simulations, we employed DUCC Hamiltonians and examined the convergence behavior of the algorithm using these DUCC-transformed Hamiltonians. 
As mentioned in the introduction, Hamiltonian downfolding carries a lot of promise for increasing the accuracy of near-term VQE calculations, while minimizing the number of qubits. 
However, all previous ADAPT-VQE convergence studies have been performed using the bare \textit{ab initio} integrals. 
In this section, we investigate the impact of dynamical correlation downfolding with DUCC on the convergence of ADAPT-VQE. 
The generalized singles and doubles operator pool was used for all simulations. The results, shown in Fig.~\ref{fig:adapt_lih_h6_h2o}, illustrate the convergence trends.

We set the convergence threshold for the norm of the gradient vector to $10^{-4}$. As observed in all simulations, ADAPT-VQE converges to the ground state of the corresponding effective Hamiltonian. This confirms that ADAPT-VQE is functioning as expected, achieving the correct energy for each case. Importantly, ADAPT-VQE takes nearly the same number of iterations to converge for the bare, A4, and A7 Hamiltonians. However, for the \ce{H2O} system, the algorithm exceeds the maximum iteration limit of 200 iterations.

This suggests that while DUCC-transformed Hamiltonians help reduce the qubit requirements and improve correlation energy within the active space, the convergence behavior of ADAPT-VQE is largely unaffected by the specific DUCC transformation. The fact that the number of iterations remains relatively the same across different Hamiltonians (Bare, A4, and A7) indicates that the transformed integrals do not significantly alter the computational complexity in terms of the number of steps required for convergence.
Importantly, we also notice no significant change in convergence behavior when including 3-body and 4-body terms in the Hamiltonian (A4(3), A7(3) and A7(34)), indicating that 2-body operator pools can still be effective when simulating more complicated downfolded effective Hamiltonians which contain higher-body operators. 
A comparison of the convergence between Bare, A4, A4(3), A7, A7(3), and A7(34) is shown in SI.

\section{Conclusions}
In this study, we explored the use of DUCC Hamiltonians, focusing on their impact on correlation energy, the influence of different amplitudes, and the convergence behavior of the ADAPT-VQE algorithm.
We demonstrated that DUCC transformations significantly improve the correlation energy, especially in bond-breaking scenarios. %, by including higher-order terms. 
The analysis also showed that CCSD amplitudes provide the most accurate DUCC energy estimates, reducing errors compared to MP2, CCD, and CCSD ($\hat T_1$=0) amplitudes, demonstrating a high sensitivity of the downfolding procedure on the chosen external amplitudes.
In terms of ADAPT-VQE convergence, we found that the algorithm successfully converges to the exact energy for Bare, A4, and A7 Hamiltonians, with similar iteration counts, highlighting that these new transformed integrals do not lead to convergence issues and allow ADAPT-VQE to recover additional correlation energy, especially early in the optimization. 
Overall, our results underscore the promise of DUCC transformations for improving quantum simulations, offering more efficient and accurate methods for quantum chemistry. 
% In future work, we aim to test DUCC with the Tensor Product Selected Configuration Interaction (TPSCI) method, which further partitions orbitals into clusters after downfolding from a large basis set to an active space. We believe this approach can improve TPSCI energies and make them more comparable to experimental data, potentially providing a more accurate and efficient method for simulating molecular systems.

\section*{Supporting Information}

\begin{acknowledgments}
 L.W.B., D.C., N.P.B., and K.K. acknowledge support from the ``Embedding Quantum Computing into Many-body Frameworks for Strongly Correlated Molecular and Materials Systems” project, which is funded by the U.S. Department of Energy (DOE), Office of Science, Office of Basic Energy Sciences, the Division of Chemical Sciences, Geosciences, and Biosciences (under FWP 72689). 
 N.P.B. also acknowledges support from the Quantum Algorithms and Architecture for Domain Science Initiative (QuAADS), a Laboratory Directed Research and Development (LDRD) Program at PNNL.
 E.B. and S.E.E. acknowledge support by the U.S. Department of Energy, Office of Science, Office of Advanced Scientific Computing Research, under Award Number DE-SC0025430.
 N.J.M acknowledges support from the U.S. Department of Energy, under Award Number DE-SC0024619. 
 The authors thank Advanced Resource Computing at Virginia Tech for use of computational resources. 
\end{acknowledgments}

\bibliography{bibliography}% Produces the bibliography via BibTeX.

\end{document}

%\preprint{APS/123-QED}

\title{Supporting Information: Qubit-efficient quantum chemistry with the ADAPT variational quantum eigensolver and double unitary downfolding}
\author{Harjeet Singh}
\thanks{These authors contributed equally to this work}
\affiliation{Department of Chemistry, Virginia Tech, Blacksburg, VA 24061, USA}
\affiliation{Virginia Tech Center for Quantum Information Science and Engineering, Blacksburg, Virginia 24061, USA}

\author{Luke W. Bertels}
\affiliation{Quantum Information Science Section, Oak Ridge National Laboratory, Oak Ridge, TN 37831, USA}
\email{bertelslw@ornl.gov}
\thanks{These authors contributed equally to this work}

\author{Daniel Claudino}
\thanks{This manuscript has been authored by UT-Battelle, LLC,under Contract DE-AC0500OR22725 with the U.S. Department of Energy. The United States Government retains and the publisher, by accepting the article for publication, acknowledges that the United States Government retains a nonexclusive, paid-up, irrevocable, worldwide license to publish or reproduce the published form of this manuscript, or allow others to do so, for the United States Government purposes. The Department of Energy will provide public access to these results of federally sponsored research in accordance with the DOE Public Access Plan.}
\affiliation{Quantum Information Science Section, Oak Ridge National Laboratory, Oak Ridge, TN 37831, USA}
\author{Sophia E. Economou}
\author{Edwin Barnes}
\affiliation{Department of Physics, Virginia Tech, Blacksburg, VA 24061, USA}
\affiliation{Virginia Tech Center for Quantum Information Science and Engineering, Blacksburg, Virginia 24061, USA}
\author{Nicholas J. Mayhall}
\email{nmayhall@vt.edu}
\affiliation{Department of Chemistry, Virginia Tech, Blacksburg, VA 24061, USA}
\affiliation{Virginia Tech Center for Quantum Information Science and Engineering, Blacksburg, Virginia 24061, USA}

\author{Nicholas P. Bauman}
\author{Karol Kowalski}
\affiliation{Physical Sciences and Computational Division, Pacific Northwest National Laboratory, Richland, WA 99354, USA}

\date{\today}

\maketitle

\tableofcontents

\section{Non-variational results from DUCC}
\begin{figure*}[!htbp]
    \centering
    \includegraphics[width=\textwidth]{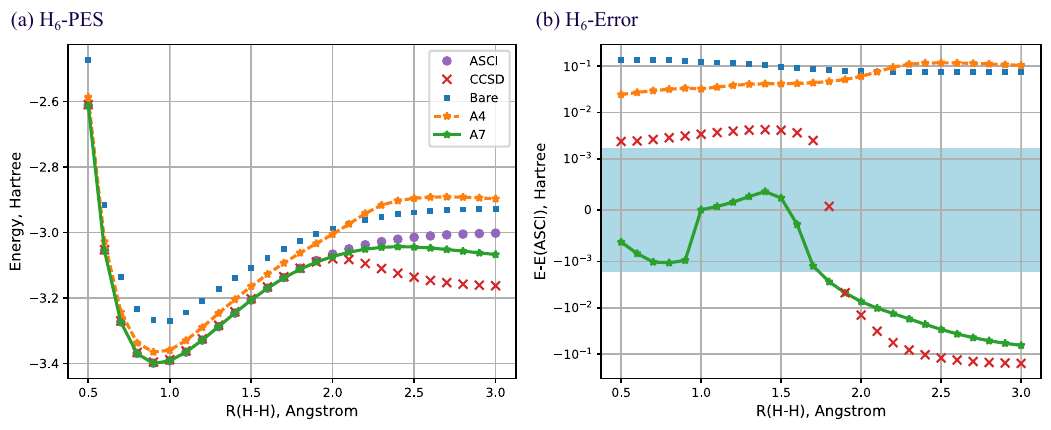}
    \caption{Comparison of the ground states of the bare, A4, and A7 Hamiltonians for symmetric linear \ce{H6} dissociation in canonical orbital basis. The left panel shows the absolute energies, while the right panel shows corresponding errors. The error for \ce{H6} is calculated relative to the extrapolated ASCI energy. The blue shaded region represents a chemical accuracy of 1kcal/mol.}
    \label{fig:pes_h6_canonical}
\end{figure*}

We performed the symmetric linear \ce{H6} dissociation in canonical orbital basis. The figure ~\ref{fig:pes_h6_canonical}a shows absolute energies and ~\ref{fig:pes_h6_canonical}b shows corresponding errors calculated relative to extrapolated ASCI energy. The A4 approximation results in energies above the bare Hamiltonian in the stretched region and A7 approximation provides non-variational energies. So, natural orbitals or natural virtual orbitals are better choice to compare to canonical orbitals for DUCC method because it provides better set of orbitals in the active space.

\section{Sensitivity of DUCC approximation with respect to active space}
\begin{figure*}[!htbp]
    \centering
    \includegraphics[width=\textwidth]{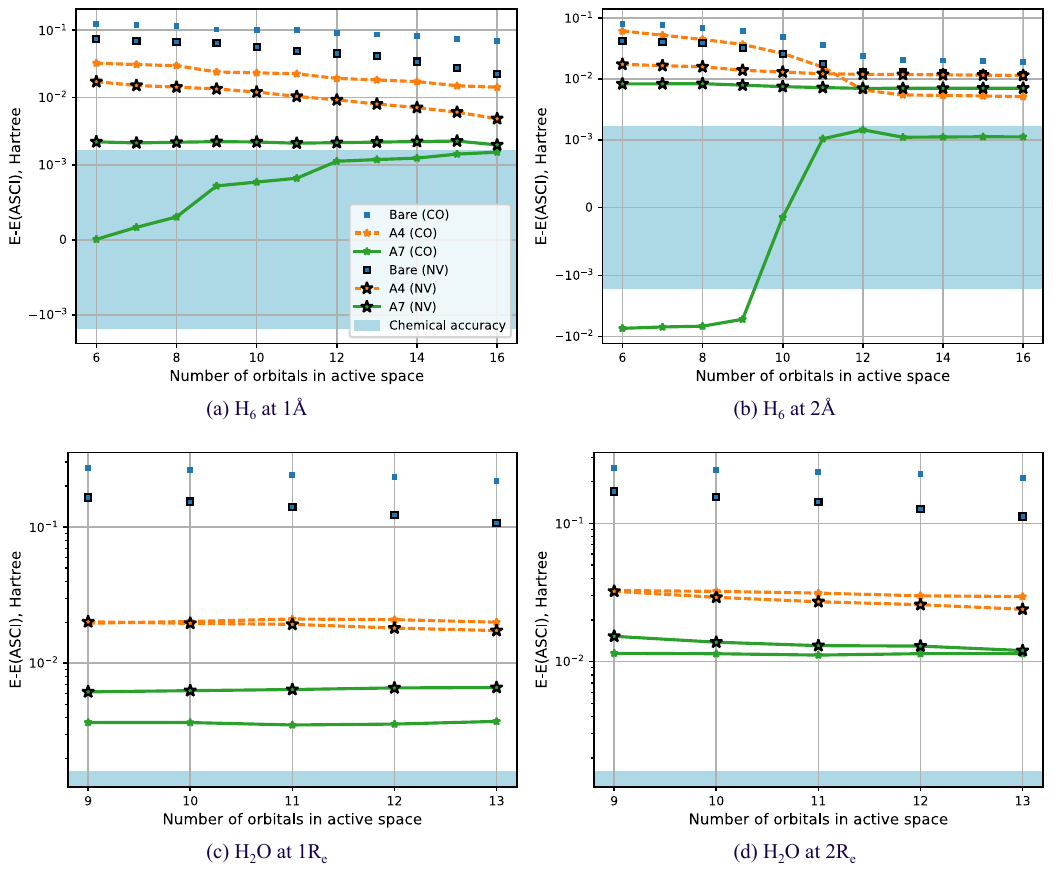}
    \caption{The errors for \ce{H6} and \ce{H2O} are shown relative to the extrapolated ASCI energy. The calculations use the cc-pVTZ basis set with different active space sizes. Here, CO refers to canonical orbitals, and NV refers to natural virtual orbitals. The blue shaded region represents a chemical accuracy of 1kcal/mol.}
    \label{fig:as_h6_h2o}
\end{figure*}

We performed DUCC calculations with various active spaces to assess whether adding more orbitals improves energies to the chemical accuracy. Calculations were performed on \ce{H6} at \ce{1\AA} and \ce{2\AA}, and on \ce{H2O} at \ce{1R_e} and \ce{2R_e}. The results, shown in Figure~\ref{fig:as_h6_h2o}, indicate that increasing the active space does not significantly change the energies. Additionally, we compared results obtained using canonical orbitals and natural virtual orbitals.

\section{Comparison of ADAPT-VQE convergence for higher body DUCC Hamiltonians}
\begin{figure*}[!htbp]
    \centering
    \includegraphics[width=\textwidth]{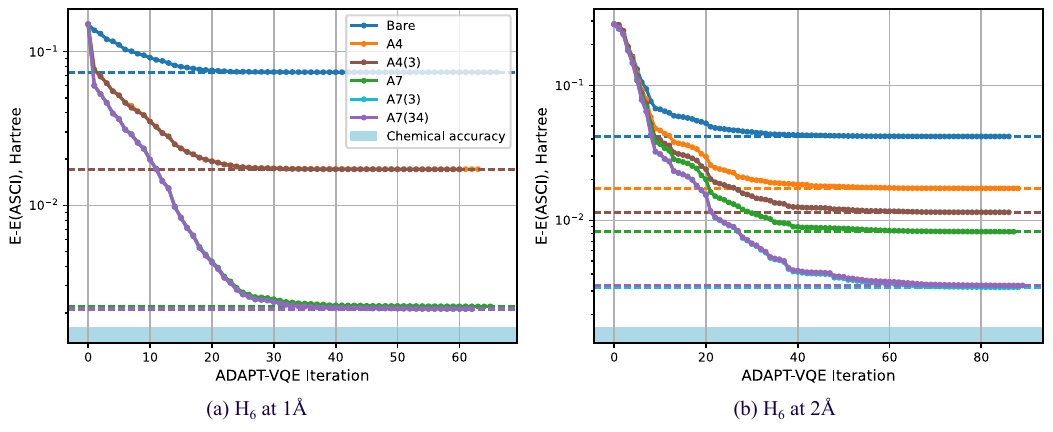}
    \caption{ Energy error versus the number of ADAPT-VQE iteration for \ce{H6} (6 orbital active space, 1 \AA ~and 2 \AA ~bond distance). Results for  bare (blue), A4 (orange), A4(3) (brown), A7 (green), A7(3) (cyan), and A7(34) (purple) Hamiltonians are shown. The dashed lines represent the exact diagonalization of the corresponding effective Hamiltonian in the active space. The blue shaded region indicates chemical accuracy (1 kcal/mol = 1.59 mHa).
}
    \label{fig:adapt_h6_higherbody}
\end{figure*}
Here we present convergence plots of ADAPT-VQE for \ce{H6} with several DUCC effective Hamiltonians, including those with contributions from three- and four-body terms. 
As in the main text, we observe that the downfolding procedure does not hinder the convergence of the algorithm even when accounting for these terms, despite only accounting for one- and two-particle excitation/de-excitations in our operator pool.

\section{Comparison of CCSD $\hat{T}_1$ and $\hat{T}_2$ amplitudes diagnostic values }
Table~\ref{tab:amps_diagnostic} summarizes the CCSD $\hat{T}_1$ and $\hat{T}_2$ diagnostics for three chemical systems—\ce{LiH}, \ce{H6}, and \ce{H2O}—evaluated at both equilibrium and stretched geometries. For each case, we report the diagnostics computed using the full set of amplitudes and external amplitudes.
\begin{table*}[!htbp]
    \caption{Diagnostic values for full and external CCSD excitation amplitudes for selected chemical systems.}
    \label{tab:amps_diagnostic}
    \centering
    \renewcommand{\arraystretch}{1.2}
    \begin{tabular*}{\textwidth}{@{\extracolsep{\fill}} l r r r r r r}
        \hline\hline
        & \ce{LiH}(1R$_e$) & \ce{LiH}(2R$_e$) & \ce{H6}(1\AA) & \ce{H6}(2\AA) & \ce{H2O}(1R$_e$) & \ce{H2O}(2R$_e$) \\
        \hline
        $\hat{T}_1$ Diagnostic & 0.007 & 0.090 & 0.014 & 0.059 & 0.006 & 0.022 \\
        $\hat{T}_{1,\text{ext}}$ Diagnostic & 0.004 & 0.035 & 0.010 & 0.058 & 0.005 & 0.020 \\
        Max $|\hat{T}_{1,\text{ext}}|$ & 0.005 & 0.049 & 0.011 & 0.075 & 0.008 & 0.046 \\
        \hline
        $\hat T_2$ Diagnostic & 0.089 & 0.193 & 0.130 & 0.450 & 0.080 & 0.140 \\
        $\hat{T}_{2,\text{ext}}$ Diagnostic & 0.015 & 0.025 & 0.071 & 0.058 & 0.053 & 0.057 \\
        Max $|\hat{T}_{2,\text{ext}}|$ & 0.011 & 0.014 & 0.024 & 0.028 & 0.022 & 0.024 \\
        \hline\hline
    \end{tabular*}
\end{table*}